\documentclass[journal,transmag]{IEEEtran}
\usepackage{braket}
\usepackage{cite}
\usepackage{hhline}
\usepackage{amsmath}
\usepackage{bm}
\usepackage{verbatim}
\usepackage{savesym}
\savesymbol{tablenum}
\usepackage{siunitx}
\usepackage{float}
\usepackage{subfigure}
\usepackage{graphicx}
\graphicspath{{./}{figures/}}

\ifCLASSINFOpdf
\else
\fi
\usepackage{geometry}                		
\usepackage{graphicx}				
\usepackage{amssymb}
\usepackage{amsmath}
\usepackage{subfig}
\usepackage{caption}
\usepackage{comment}

\usepackage[utf8]{inputenc}
\usepackage{amsmath}
\numberwithin{equation}{section}
\begin{document}
%
\title{Casimir energy with chiral fermions \\
on a quantum computer}


\author{\IEEEauthorblockN{Juliette K. Stecenko\IEEEauthorrefmark{1}, Yuan Feng\IEEEauthorrefmark{2},Michael McGuigan\IEEEauthorrefmark{3}}
\IEEEauthorblockA{\IEEEauthorrefmark{1}Rutgers University (Current affiliation Dept. of Physics, Univ. Connecticut)}
\IEEEauthorblockA{\IEEEauthorrefmark{2}Pasadena City College}
\IEEEauthorblockA{\IEEEauthorrefmark{3} Brookhaven National Laboratory}
\thanks{}}


%



\IEEEtitleabstractindextext{%
\begin{abstract}
In this paper we discuss the computation of Casimir energy on a quantum computer. The Casimir energy is an ideal quantity to calculate on a quantum computer as near term hybrid classical quantum algorithms exist to calculate the ground state energy and the Casimir energy gives  physical implications for this quantity in a variety of settings.  Depending on boundary conditions and whether the field is bosonic or fermionic we illustrate how the Casimir energy calculation can be set up on a quantum computer and calculated using the Variational Quantum Eigensolver algorithm with IBM QISKit. We compare the results based on a lattice regularization with a finite number of qubits with the continuum calculation for free boson fields, free fermion fields and chiral fermion fields. We use a regularization method introduced by Bergman and Thorn to compute the Casimir energy of a chiral fermion. We show how the accuracy of the calculation varies with the number of qubits. We show how the number of Pauli terms which are used to represent the Hamiltonian on a quantum computer scales with the number of qubits. We discuss the application of the Casimir calculations on quantum computers to cosmology,  nanomaterials, string models, Kaluza Klein models and dark energy.
 \\
\end{abstract}

}

\maketitle

\IEEEdisplaynontitleabstractindextext

%
\IEEEpeerreviewmaketitle

\section{Introduction}

In this paper we study the Casmir energy of a quantum physical system of bosons and fermions s which has applications to nanoscience, early Universe cosmology and Kaluza Klein theory and String theory.
\cite{Zeldovich:1984vk}
\cite{Milton:2014tca}
\cite{Brevik:2000zb}
\cite{Milton:2002hx}
\cite{Appelquist:1982zs}
\cite{Candelas:1983ae}
\cite{Kantowski:1987ct}
\cite{Ordonez:1985kz}
\cite{Accetta:1986vq}
\cite{Gleiser:1987mg}
\cite{Bailin:1985we}
\cite{Bailin:1987jd}
\cite{Bailin:1984xf}
\cite{Myers:1985uy}
\cite{Birmingham:1988iv}
\cite{Camporesi:1990wm}
\cite{Myers:1987rp}
\cite{Milton:2001yy}
\cite{Yan:1987dw}
\cite{Cho:2004ew}
\cite{Stuckey:1990te}
\cite{RandjbarDaemi:1983jz}
\cite{Dowker:1990br}
\cite{Brink:1986ja}
\cite{Thorn:1988em}
\cite{Ginsparg:1986wr}
\cite{Nair:1986zn} We study the simplest system of boson and fermion fields in one time and one space dimension but our methods can be extended to more realistic systems. We use the Variational Quantum Eigensolver (VQE) algorithm for our calculations which can be efficiently run on current quantum computers such as the IBM Q \cite{Kandela}. We use the QISKit application programming interface (API) to run the code on a quantum simulator with various number of qubits depending on the problem size. We are able to get accurate results for the Casimir energy using these techniques. Our results indicate a growing number of Pauli terms are required to represent the problem on a quantum computer as the lattice size grows.   Parallel quantum computing in which one performs calculations by summing over  computation on smaller partitions each of which uses a smaller number of Pauli terms may make larger calculations  possible in the future.

This study will focus on the application of quantum computing to the field of cosmology. We will demonstrate how one may use quantum computing to calculate the Casimir energy of scalar fields consisting of particles of integer spin (Bosons) and particles of half-integer spin (Fermions). Interactions on small scales, such as the Casimir effect, are applicable to cosmology when we consider an extremely early compact Universe as Quantum cosmology concentrates on the Universe extremely close to the Big Bang. 

Quantum computing makes use of quantum superposition (the additive property of quantum states), quantum interference (the destructive property of quantum states) and quantum entanglement  to perform complex calculations. The quantum states acted upon by these phenomena scale exponentially with the number of quantum bits, qubits, these differ from classical bits, which are represented by 1s and 0s, by being able to represent an infinite amount of state combinations. Initial  states represented by qubits are  transformed by Unitary operations and computations are performed using quantum logic gates, which operate on the  qubits whose final state undergoes measurement. In this study we use IBM's Qiskit, an open-source tool made to utilize quantum computing through python. We used these tools to run on IBM's Quantum simulator, being able to run a maximum of 32 qubits. These tools will be utilized through a variational quantum eigensolver (VQE). The VQE can be used to find ground state eigenvalues of the Hamiltonians, by repeating the quantum calculation for an appropriate optimization steps. This study had the secondary goals of minimizing the number of qubits and Paulis needed to do complex calculations. \\
\par Additionally our runs will be using the COBLYA, Constrained Optimization BY Linear Approximation, which is based in the SciPy Python library's optimization class. COBLYA constructs linear polynomial approximations to the input by interpolation at vertices, this repeats to yield new optimized data points. We also use the Sequential Least SQuares Programming (SLSQP) optimizer. These optimizers are included in Qiskit's Aqua library.

\section{Casimir Effect}
The Casimir Effect describes the physical forces or changes that come as a result of quantum fields together with their boundary conditions, we can describe this effect in terms of the Casimir energy. Casimir energy was first proposed to explain  that the interaction energy between a neutral atom and perfect conducting plate would result in a reduction in energy, that is inversely proportional to the distance of the two bodies. Thus the energy must rely on both the quantum field itself and the boundary conditions. We focused on  fields describing Bosons and Fermions. These fields can have four boundary conditions, as follows, where $a$ is the radius of the spatial dimension and $\epsilon $ the Casimir energy density \cite{Birrell:1982ix}.

\noindent Boson with periodic boundary conditions:
\begin{equation}
\epsilon= -\frac{\pi}{6 a^2}
\end{equation}
Twisted Boson with antiperiodic boundary condition:
\begin{equation}
\epsilon= \frac{\pi}{12 a^2}
\end{equation}
Fermion with periodic boundary conditions:
\begin{equation}
\epsilon= 4\frac{\pi}{6 a^2}
\end{equation}
Twisted Fermion with antiperiodic boundary condition:
\begin{equation}
\epsilon= -4\frac{\pi}{12 a^2}
\end{equation}

\section{Representation in terms of qubits}
\par In this study we focused on two aspects, creating accurate Hamiltonians and running these Hamiltonians in a variational quantum eigensolver (VQE). We begin by defining our creation and annihilation matrices: 
\begin{equation}
    {a_i} = {I_1} \otimes  \ldots {I_{i - 1}} \otimes \left( {\begin{array}{*{20}{c}}
0&1&0&0\\
0&0&{\sqrt 2 }&0\\
0&0&0&{\sqrt 3 }\\
0&0&0&0
\end{array}} \right) \otimes {I_{1}} \otimes  \ldots {I_{n-i}}
\label{eqn:aop}
\end{equation}
\begin{equation}
    {c_i} = {\sigma_1} \otimes  \ldots {\sigma_{i - 1}} \otimes \left( {\begin{array}{*{20}{c}}
0&1\\
0&0
\end{array}} \right) \otimes {I_{1}} \otimes  \ldots {I_{n-i}}
\label{eqn:cop}
\end{equation}
For bosons the operators \ref{eqn:aop} and \ref{eqn:cop} act as ladder operators, being lower and raising operators respectively, and obey commutation relations while in Fermions they obey  anticommutation relations. \\ 
\par As stated prior, our we chose to use scalar fields of Bosons and Fermions in the form of scalar rings to represent our system. We find each boundary condition has a particular set of normal mode frequencies, periodic Bosons as in equation \ref{eqn:PBfreq}, twisted Bosons in equation \ref{eqn:TBfreq}, periodic Fermions in equation \ref{eqn:PFfreq}, twisted Fermions in equation \ref{eqn:TFfreq}. \\

\begin{equation}
    \omega_{BP_i}(i,N)=\frac{8}{2N+1}2\sin(\frac{2\pi i}{4N+2})
    \label{eqn:PBfreq}
\end{equation}

\begin{equation}
    \omega_{BT_i}(i,N)=\frac{8}{2N+1}2\sin(\frac{2\pi(i+\frac{1}{2})}{4N+2})
    \label{eqn:TBfreq}
\end{equation}
\begin{equation}
    \omega_{FP_i}(i,N)=\frac{32}{2N+1}2\sin(\frac{2\pi}{4N+2})
    \label{eqn:PFfreq}
\end{equation} 
\begin{equation}
    \omega_{FT_i}(i,N)=\frac{32}{2N+1}2\sin(\frac{2\pi(i+\frac{1}{2})}{4N+2})
    \label{eqn:TFfreq}
\end{equation}
	
	 These will then factor into the Hamiltonians, which are dependent on the spin of the particles. 
\begin{equation}
    H_{B_i}=\omega_{B_i}(a_i^Ta_i + \frac{1}{2}I)
    \label{eqn:hamb}
\end{equation}
\begin{equation}
  H_{F_i}=\omega_{F_i}(c_i^Tc_i - \frac{1}{2}I)
    \label{eqn:hamf}
\end{equation}
\par Additionally we add a correction factor to these ground state energy results, these correction factors correspond to the a substraction term which yields the correct Casimir energy in the continuum limit. Particularly this also depends of the particles present and for Bosons the corrections is:
\begin{equation}
\epsilon_C= -\frac{8}{\pi}
\end{equation}
For Fermions:
\begin{equation}
\epsilon_C=+\frac{32}{\pi}
\end{equation}
For both Fermions and Bosons:
\begin{equation}
\epsilon_C= +\frac{24}{\pi}
\end{equation}
We use Mathematica software to create these Hamiltonians  for number of  lattice sites from $n=1$ to $n=8$ for each respective boundary condition. These are output as text files and input into the VQE using Qiskit software. The VQE will solve for the ground state energy of the system, using the  equation \ref{eqn:vqe}.
\begin{equation}
    E_0 \leq \bra{\psi}H\ket{\psi}\equiv \braket{H}
    \label{eqn:vqe}
\end{equation}

\section{Results of 1+1 Casimir Energy quantum computation}

\begin{table}[H]
\begin{tabular}{rrrr}
\multicolumn{1}{l}{\textbf{n}} & \multicolumn{1}{l}{\textbf{Exact Energy}} & \multicolumn{1}{l}{\textbf{VQE Energy}} & \multicolumn{1}{l}{\textbf{Percent Difference}} \\
1 & -0.2371 & -0.2371 & -6.13E-06 \\
2 & -0.0843 & -0.0843 & -7.11E-05 \\
3 & -0.0429 & -0.0428 & -2.52E-01 \\
4 & -0.0259 & -0.0259 & -1.42E-04 \\
5 & -0.0173 & -0.0172 & -6.72E-01 \\
6 & -0.0124 & -0.0124 & -7.02E-04 \\
7 & -0.0093 & -0.0093 & -2.18E-02 \\
8 & -0.0073 & -0.0073 & -1.53E-03
\end{tabular}
\caption{Periodic Bosons ground state energies with correction.}
\label{tab:pb}
\end{table}
\begin{table}[H]
\begin{tabular}{rrrr}
\multicolumn{1}{l}{\textbf{n}} & \multicolumn{1}{l}{\textbf{Exact Energy}} & \multicolumn{1}{l}{\textbf{VQE Energy}} & \multicolumn{1}{l}{\textbf{Percent Difference}} \\
1 & 0.1202 & 0.1202 & 4.05E-05 \\
2 & 0.3479 & 0.3482 & 6.27E-02 \\
3 & 0.3386 & 0.3388 & 4.51E-02 \\
4 & 0.3031 & 0.3033 & 6.31E-02 \\
5 & 0.2688 & 0.2689 & 4.82E-02 \\
6 & 0.2397 & 0.2397 & 3.45E-05 \\
7 & 0.2156 & 0.2156 & 1.32E-03 \\
8 & 0.1955 & 0.1955 & 6.46E-05
\end{tabular}
\caption{Twisted Boson ground state energies with correction.}
\label{tab:tb}
\end{table}
\begin{table}[H]
\begin{tabular}{rrrr}
\multicolumn{1}{l}{\textbf{n}} & \multicolumn{1}{l}{\textbf{Exact Energy}} & \multicolumn{1}{l}{\textbf{VQE Energy}} & \multicolumn{1}{l}{\textbf{Percent Difference}} \\
1 & 0.9483 & 0.9483 & 2.42E-06 \\
2 & 0.3373 & 0.3373 & 2.41E-05 \\
3 & 0.1715 & 0.1715 & 6.72E-06 \\
4 & 0.1036 & 0.1036 & 7.69E-05 \\
5 & 0.0693 & 0.0693 & 7.11E-05 \\
6 & 0.0496 & 0.0496 & 2.60E-04 \\
7 & 0.0373 & 0.0373 & 1.04E-04 \\
8 & 0.029 & 0.029 & 7.41E-05
\end{tabular}
\caption{Periodic Fermion ground state energies with correction.}
\label{tab:pf}
\end{table}
\begin{table}[H]
\begin{tabular}{rrrr}
\multicolumn{1}{l}{\textbf{n}} & \multicolumn{1}{l}{\textbf{Exact Energy}} & \multicolumn{1}{l}{\textbf{VQE Energy}} & \multicolumn{1}{l}{\textbf{Percent Difference}} \\
1 & -0.4808 & -0.4808 & -8.79E-07 \\
2 & -1.3918 & -1.3918 & -3.02E-06 \\
3 & -1.3545 & -1.3545 & -2.24E-06 \\
4 & -1.2123 & -1.2123 & -6.09E-06 \\
5 & -1.0752 & -1.0752 & -3.32E-06 \\
6 & -0.9589 & -0.9589 & -9.03E-06 \\
7 & -0.8623 & -0.8623 & -2.97E-06 \\
8 & -0.782 & -0.782 & -2.91E-06
\end{tabular}
\caption{Twisted Fermion ground state energies with correction}
\label{tab:tf}
\end{table}
Upon calculating the ground state energies, we find a -4 times difference from the ground state energies of Fermions to Bosons between respective frequency types and number of lattice sites for both the exact and VQE calculations, this is expected. Additionally, we find a remarkably minimal percent difference between the exact and VQE eigenvalues. This is far more accurate than initially expected.  
\begin{figure}[H]
    \centering
    \includegraphics[scale=0.15]{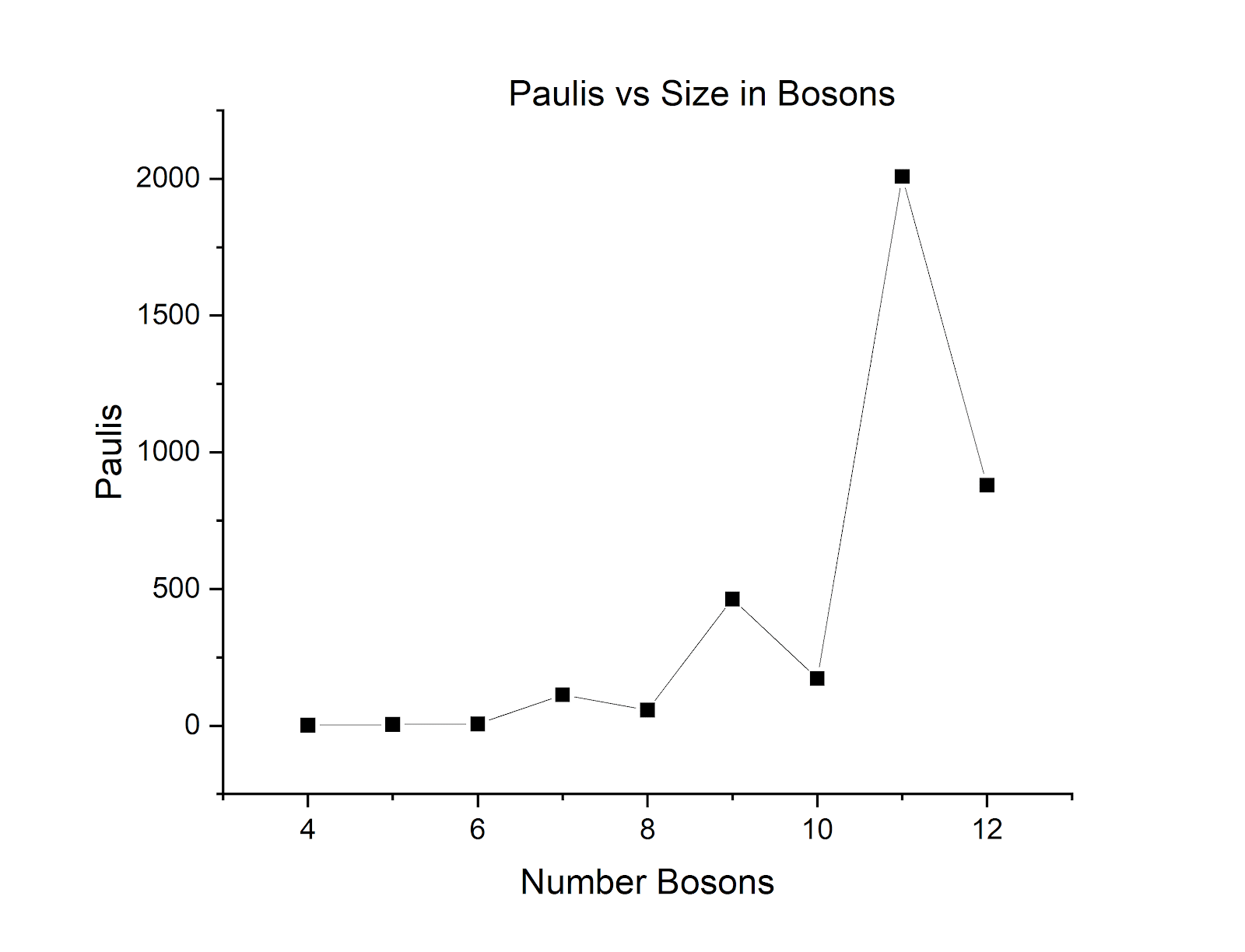}
    \caption{Paulis used to number of Bosons in the ring structure.}
    \label{fig:Pauli}
\end{figure}

\par During our runs we found that  larger Hamiltonians were realized with an unusually large amount of Paulis, as seen in figure \ref{fig:Pauli}, thus we chose to cap the amount of lattice sites to $n=8$. Additionally we focused on optimizing the Hamiltonians to allow to minimize run time. Thus the number of optimization steps and shots varies with the size of the Hamiltonians. While ideally more steps and shots may lead to better accuracy  they may also increase the run time. Our runs as shown in figure \ref{fig:Pauli}. We were able to run $n=8$ Bosons for both periodic and twisted, and $n=8$ Fermions for both periodic and twisted, for 4 Qubits and 5 Paulis. The convergence of our optimizers, with subtractions, can be seen in figures \ref{fig:PB},\ref{fig:TB},\ref{fig:PF}, and \ref{fig:TF}.
\begin{figure}[H]
    \centering
    \includegraphics[scale=0.3]{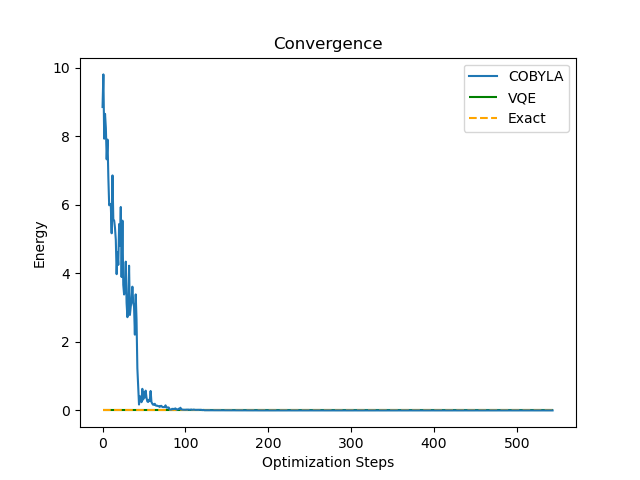}
    \caption{Convergence of the optimizer for $n=8$ periodic Bosons, with correction.}
    \label{fig:PB}
\end{figure}
\begin{figure}[H]
    \centering
    \includegraphics[scale=0.3]{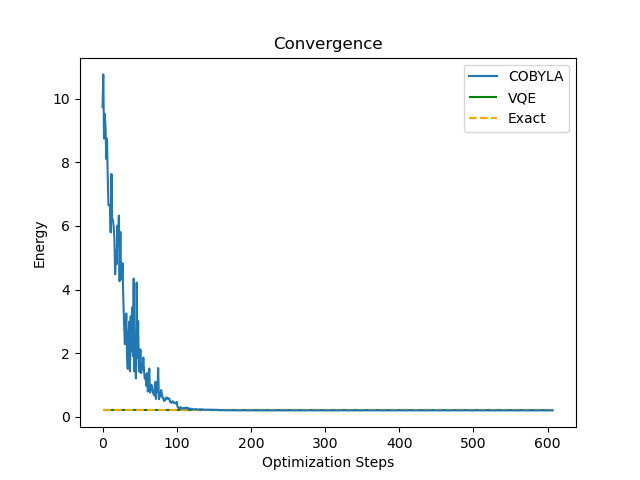}
    \caption{Convergence of the optimizer for $n=8$ twisted Bosons, with correction.}
    \label{fig:TB}
\end{figure}
\begin{figure}[H]
    \centering
    \includegraphics[scale=0.3]{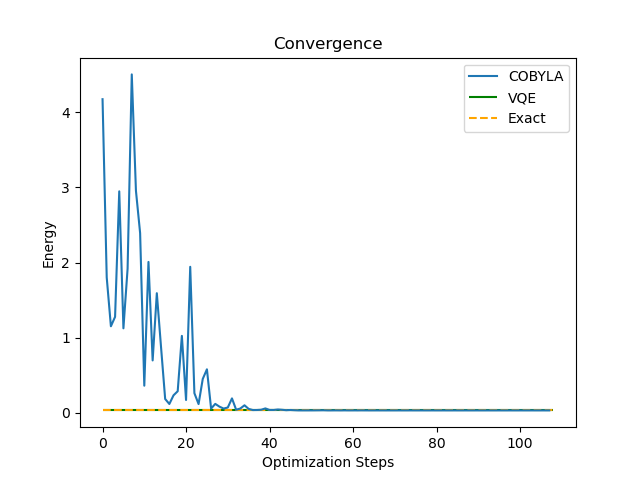}
    \caption{Convergence of the optimizer for $n=8$ periodic Fermions,with correction.}
    \label{fig:PF}
\end{figure}
\begin{figure}[H]
    \centering
    \includegraphics[scale=0.3]{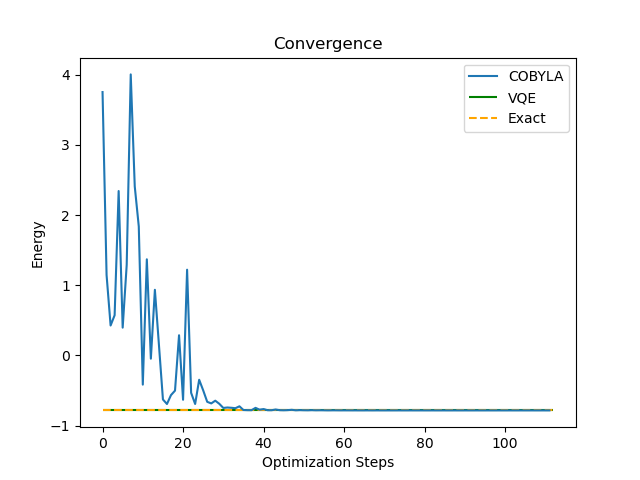}
    \caption{Convergence of the optimizer for $n=8$ twisted Fermions, with correction.}
    \label{fig:TF}
\end{figure}

\section{Casimir Energy calculation on a lattice}

Using a lattice approximation to the one plus one dimensional quantum fields the mode frequencies of the fields can be derived from Fourier decomposition.
T
The Casimir energy can then be computed in the lattice regularization by summing over positive node frequencies and subtracting off  a finite piece. For periodic bosons this is:
\begin{equation}\varepsilon  = \frac{1}{2}\sum\limits_{n = 0}^{N - 1} {{\omega _n}}  - {\delta _{per}}\end{equation}
with $\delta_{per} = \frac{8}{\pi}$. For large values of $N$ this is approximately:
\begin{equation}\varepsilon  =  - \frac{\pi }{{6{N^2}}} + \frac{\pi }{{6{N^3}}} - \frac{{180\pi  + {\pi ^3}}}{{1440{N^4}}} +  \ldots \end{equation}
which agrees with the continuum result for large number of lattice sites $N$.
For the periodic fermions we have:
\begin{equation}\varepsilon  = -\frac{1}{2}\sum\limits_{n = 0}^{N - 1} {{4\omega _n}}  + {4 \delta _{per}}\end{equation}
For large values of $N$ this is approximately:
\begin{equation}\varepsilon  =   \frac{4\pi }{{6{N^2}}} - \frac{4 \pi }{{6{N^3}}} + 4\frac{{180\pi  + {\pi ^3}}}{{1440{N^4}}} +  \ldots \end{equation}
which agrees with the continuum result for large number of lattice sites $N$.
For twisted bosons we have:
\begin{equation}\varepsilon  = \frac{1}{2}\sum\limits_{n = 0}^{N - 1} {{\tilde \omega _n}}  - { \delta _{per}}\end{equation}
For large values of $N$ this is approximately:
\begin{equation}\varepsilon  = \frac{\pi }{{12{N^2}}} - \frac{\pi }{{12{N^3}}} + \left( {\frac{\pi }{{16}} - \frac{{7{\pi ^3}}}{{11520}}} \right)\frac{1}{{{N^4}}} +  \ldots \end{equation}
which agrees with the continuum result for large number of lattice sites $N$.
Finally for  twisted fermions we have:
\begin{equation}\varepsilon  = -\frac{1}{2}\sum\limits_{n = 0}^{N - 1} {{4 \tilde \omega _n}}  + { 4\delta _{per}}\end{equation}
For large values of $N$ this is approximately:
\begin{equation}\varepsilon  = -\frac{4\pi }{{12{N^2}}} + \frac{4 \pi }{{12{N^3}}} -4 \left( {\frac{\pi }{{16}} - \frac{{7{\pi ^3}}}{{11520}}} \right)\frac{1}{{{N^4}}} +  \ldots \end{equation}
which agrees with the continuum result for large number of lattice sites $N$.
The calculation in terms of qubits can then be realized by expanding the Hamiltonian as a expansion of Pauli terms of the form ${I_1} \otimes {X_2} \otimes {Y_3} \otimes  \ldots {Z_N}$ etc with coefficients determined by taking traces or directly from the QISKit software.




\section{Casimir energy of a chiral fermion}

It is difficult to regularize a chiral fermion computation. For example the theorem of Nielsen-Ninomiya \cite{Nielsen:1980rz}
\cite{Wilczek:1987kw} forbids a lattice regularization of a chiral fermion. Nevertheless Bergman and Thorn \cite{Bergman:1995wh} devised a method to treat a chiral fermion in 1+1 dimensions using a modification of a Wilson regulated fermion. They applied their technique to the lattice regularization of the world sheet Hamiltonian heterotic string but the same approach can be used for other 1+1 dimensional fermionic systems. We will apply their method in this paper to calculate the Casimir energy of a chrial fermion on a quantum computer.

A Wilson regulated fermion in 1+1 dimensions \cite{Zache:2018jbt} can be described by the following Hamiltonian 0f Creutz and Horvath \cite{Creutz:1994ny} 
\[H = i\sum\limits_j {\left( {c_{j + 1}^\dag {c_j}} \right)}  - i\sum\limits_j {\left( {\tilde c_{j + 1}^\dag {{\tilde c}_j}} \right)}  + \]
\begin{equation}\sum\limits_j {\left( {2c_j^\dag {{\tilde c}_j} - c_j^\dag {{\tilde c}_{j + 1}} - c_{j + 1}^\dag {{\tilde c}_j}} \right)}  + c.c.
\end{equation}
where the fermion field is $ \psi  = \left( \begin{array}{l}
c\\
{\tilde c}
\end{array} \right)$. The Energy Dispersion relation from this Hamiltonian is plotted in figure 6, The relation is chirally symmetric with the same behavior for left and right moving fermions. The mixing between left and right fermions comes from the Wilson term which is the third term in the above Hamiltonian and has the effect of lifting the doubled fermion states that would be present if the Hamiltonin just consisted of the first two terms.

Bergman and Thorn modified this Hamiltonian by introducing a parameter $\eta$ so the the Hamiltonian is:
\[H = i\sum\limits_j {\left( {c_{j + 1}^\dag {c_j}} \right)}  - i\eta \sum\limits_j {\left( {\tilde c_{j + 1}^\dag {{\tilde c}_j}} \right)}  + \]
\begin{equation}\sum\limits_j {\left( {2c_j^\dag {{\tilde c}_j} - c_j^\dag {{\tilde c}_{j + 1}} - c_{j + 1}^\dag {{\tilde c}_j}} \right)}  + c.c.
\end{equation}
The quadratic Hamiltonian is of the form \begin{equation}H = ({c^\dag },{{\tilde c}^\dag })[t]\left( \begin{array}{l}
c\\
{\tilde c}
\end{array} \right)\end{equation}
where $t$ is a Hertmition matrix given by:
\begin{equation}t = \left( {\begin{array}{*{20}{c}}
A&B\\
{{B^\dag }}&{ - \eta A}
\end{array}} \right)\end{equation}
where
\begin{equation}A = \left( {\begin{array}{*{20}{c}}
0&{ - i}&0&0&0&i\\
i&0&{ - i}&0&0&0\\
0&i&0&{ - i}&0&0\\
0&0&i&0&{ - i}&0\\
0&0&0&i&0&{ - i}\\
{ - i}&0&0&0&i&0
\end{array}} \right)\end{equation}
and 
\begin{equation}B = \left( {\begin{array}{*{20}{c}}
2&{ - 1}&0&0&0&{ - 1}\\
{ - 1}&2&{ - 1}&0&0&0\\
0&{ - 1}&2&{ - 1}&0&0\\
0&0&{ - 1}&2&{ - 1}&0\\
0&0&0&{ - 1}&2&{ - 1}\\
{ - 1}&0&0&0&{ - 1}&2
\end{array}} \right)\end{equation}
The dispersion relation associated with the $t$ matrix is show in figure 6 for the ordinary Wilson fermion with $\eta=1$ and in figure 7 for $\eta=5$. We can see that the two left moving modes are raised in energy and are effectively gapped from the system for $eta=5$.
\begin{figure}[H]
    \centering
    \includegraphics[scale=0.5]{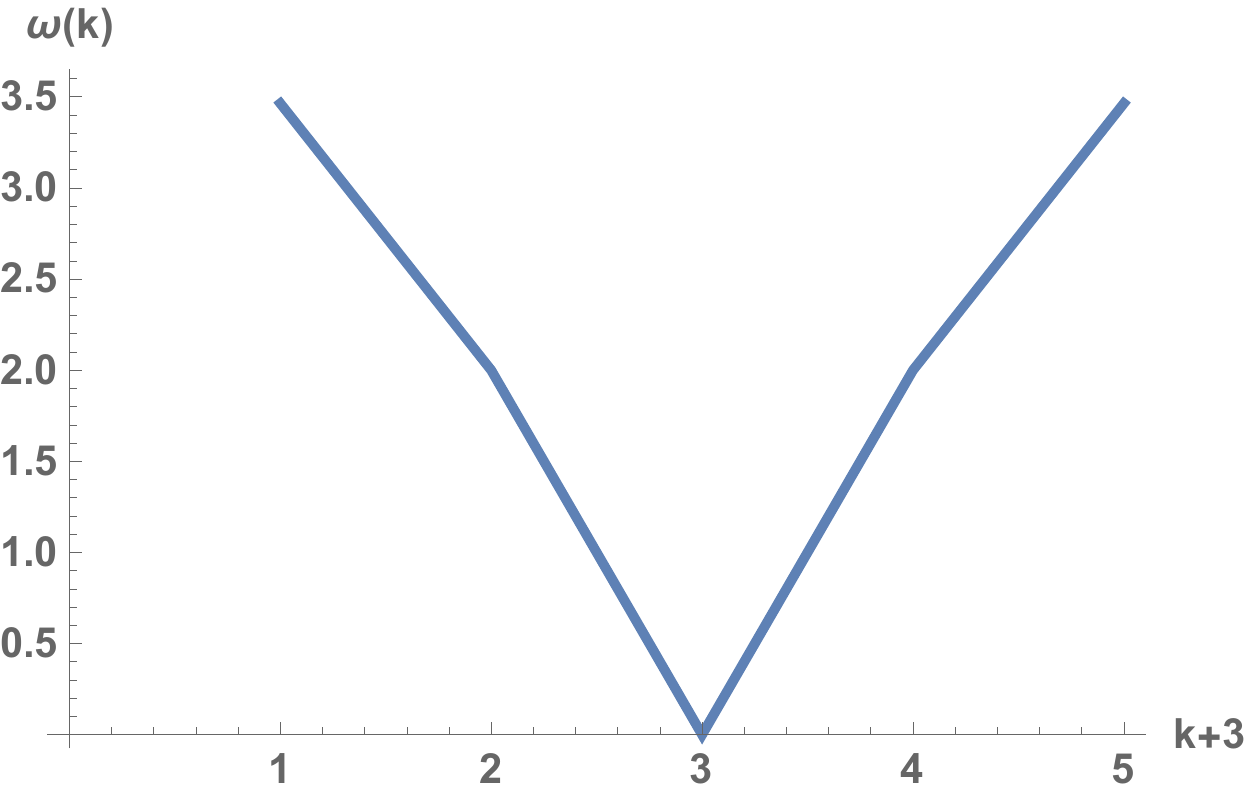}
    \caption{Dispersion relation for non chiral Wilson fermion with $   \eta=1$. Note the left movers are symmetric with the right movers}
    \label{fig:PB}
\end{figure}
\begin{figure}[H]
    \centering
    \includegraphics[scale=0.5]{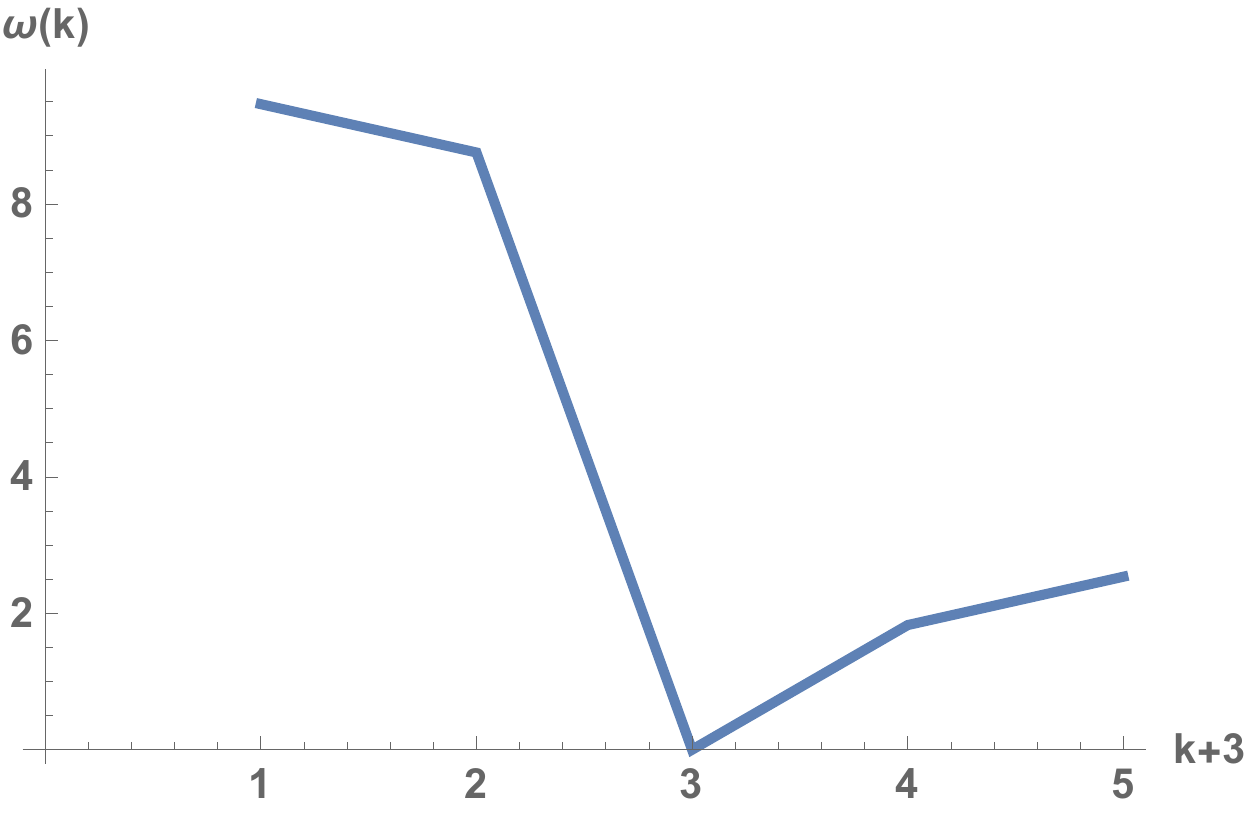}
    \caption{Dispersion relation for chiral fermion with $ \eta =5 $. Note the left movers have been raised in energy above the maximum energy of the right movers}
    \label{fig:PB}
\end{figure}
\begin{figure}[H]
    \centering
    \includegraphics[scale=0.4]{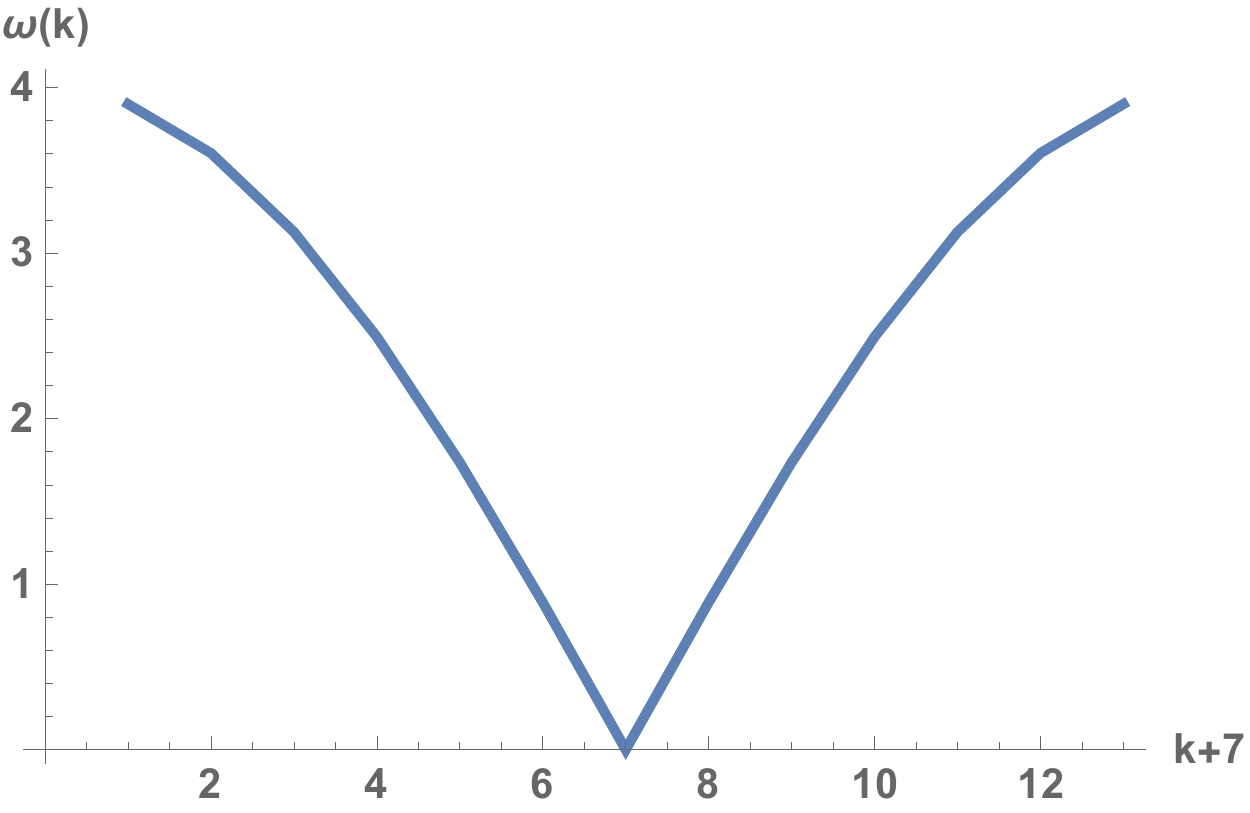}
    \caption{Dispersion relation for non chiral Wilson fermion with $   \eta=1$ with six left moving and six right moving fermion modes. Note the left movers are symmetric with the right movers}
    \label{fig:PB}
\end{figure}
\begin{figure}[H]
    \centering
    \includegraphics[scale=0.4]{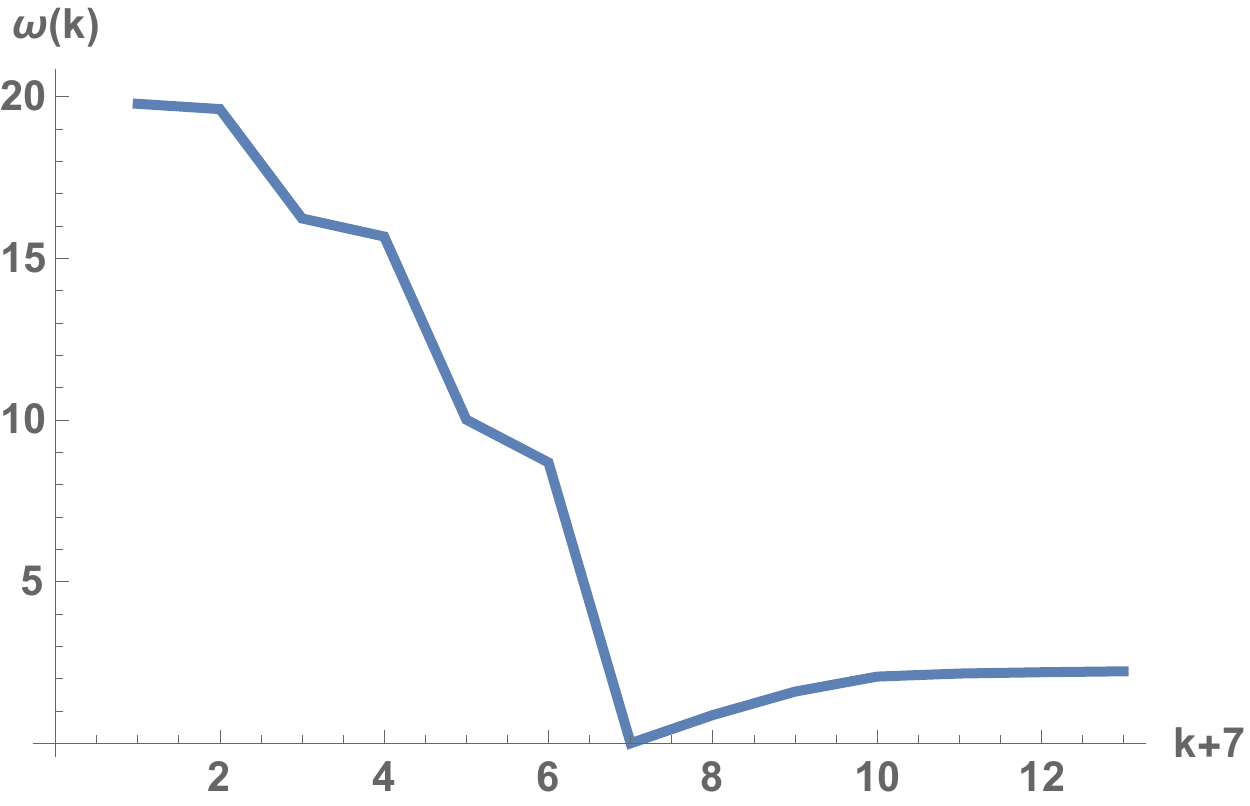}
    \caption{Dispersion relation for chiral fermion with $ \eta =10 $ and six left moving and six right moving fermion modes. Note the left movers have been raised in energy above the maximum energy of the right movers}
    \label{fig:PB}
\end{figure}
\begin{figure}[H]
    \centering
    \includegraphics[scale=0.25]{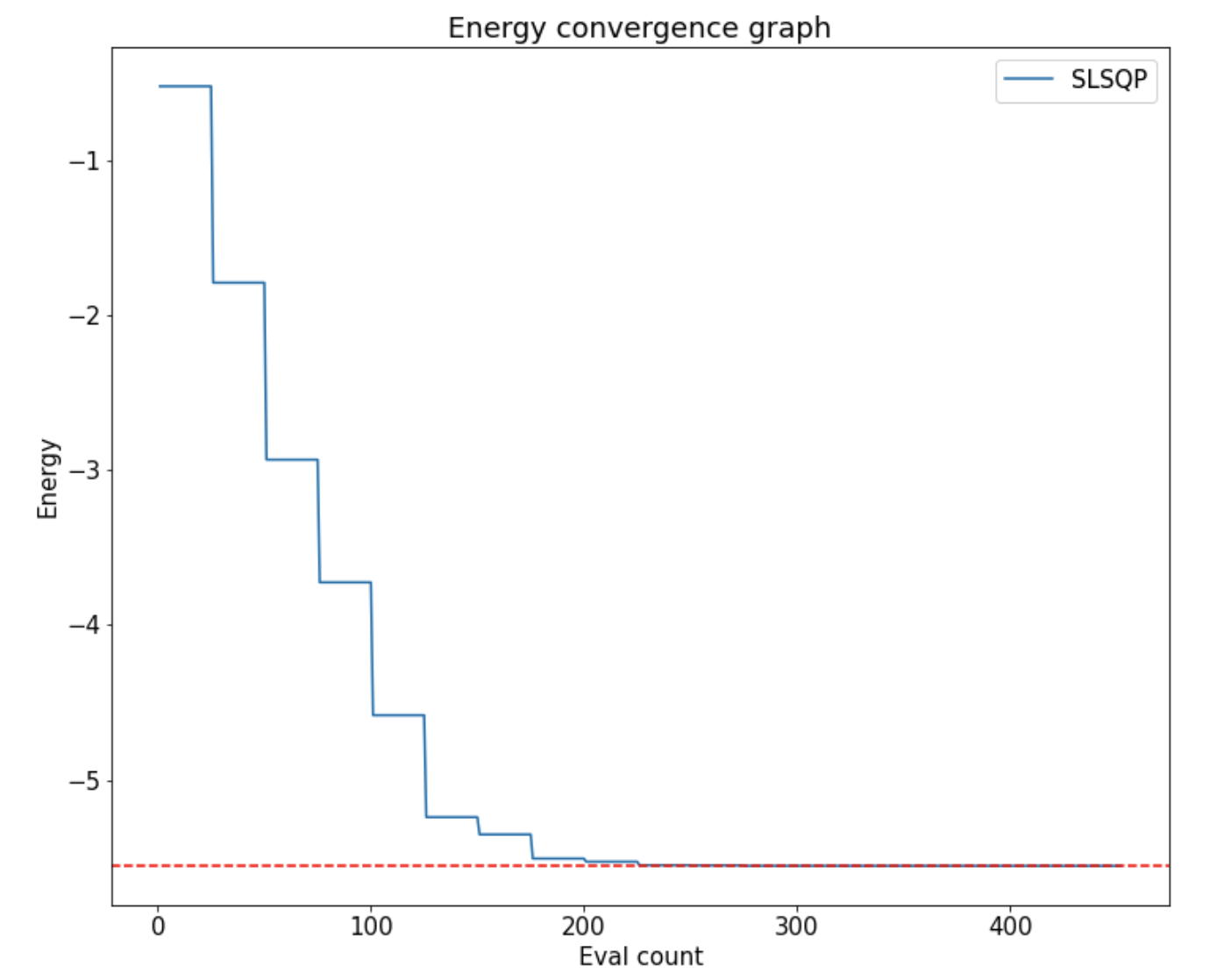}
    \caption{Convergence plot for quantum computation of the ground state energy of a chiral fermion with six right moving modes using the Bergman-Thorn scheme with $\eta = 10$. }
    \label{fig:PB}
\end{figure}
As described by Bergman and Thorn \cite{Bergman:1995wh}  the dispersion relation is modified so that the left movers require much great energy to excite their modes. In particular on a lattice even the lowest frequency left moving modes can be raised in energy and the left movers are effectively gapped from the system. We can apply their scheme for chiral fermions to the calculation of Casimir energy on a quantum computer. By considering large $t$  matrices one can describe larger fermion system. For a $t$ matrix that is $28\time 28$ we can describe a system with six right moving fermions and six left moving fermions whose dispersion relation is shown in figure 9 with  $\eta = 10$.  Performing the calculation using IBM QISKit and the VQE algorithm with the Sequential Least SQuares Programming (SLSQP) optimizer we find a ground state energy $ E_0 =-5.55433399$ which is in close agreement with the exact value of $E_{exact0}=-5.55433587$
The subtraction of a constant term from the ground state energy  calculation will yield the Casimir energy. In the Thorn Bergman case the subtraction term will be $\eta$ dependent unlike the case of a non-chiral or left right symmetric fermion computation. Performing this subtraction for $\eta = 10$ of $-5.57571769$ we find that the Casimir energy from the six right moving fermions is $ E_C=0.0213837$ which is in good agreement with the exact expression for the Casimir energy of a chiral fermion which is $E_{exactC} = 2\frac{\pi}{6 (7)^2} = 0.0213714$.

One difficulty of the Bergman-Thorn computation of Casimir energy is the the left movers will also get a large value of Casimir energy. This would dominate the the contribution of the right movers and would need to be subtracted away by adding a constant term so that the mass relations of the heterotic string on the lattice would yield the correct mass squared relations. A possible resolution would be to use a 2+1 dimensional lattice fermion action as in Creutz-Horvath \cite{Creutz:1994ny}. Another issue to be investigated is the effect of interactions on the Bergman-Thorn approach to chiral fermions.

In the context of String/M-theory the discrete chiral heterotic world sheet action is related to a lattice regularized world volume action of M-theory which reduces to the heterotic string by dimensional reduction. Interestingly this world-volume to chiral world-sheet dimensional reduction is similar to the Horova-Witten dimensional reduction of M-theory itself \cite{Horava:1995qa} to the heterotic effective action which is also chiral.

\section{Conclusions}


In this paper we applied quantum computation to the calculation of Casimir energy. This quantity has important applications to nanoscience, cosmology and Kaluza Klein Theory. We used the Variational Quantum Eigensolver algorithm (VQE) which can efficiently run on existing quantum computers such as the IBM Q with up to 60 qubits. Our results indicate to we can accurately compute the Casimir energy for boson and fermion systems for up the eight lattice sites. We also introduced the notion of parallel quantum computation that allows the Casimir boson and fermion systems of larger lattice size to be calculated on multiple quantum computers or partitions of a single quantum computer. We also analyzed the number of Pauli terms required to represent the Hamiltonian of the system on the quantum computer and studied how this increased with problem size. Parallel quantum computation on future quantum computers will  allow one to work with with smaller number of Pauli terms on each partition to achieve even more accurate computations of Casimir energy.

We were able to compute the Casimir energy for 1+1 dimensional cosmologies in the forms of Bosons and Fermions in rings with up to $n=8$ lattice sites, using quantum computing. We did this through using a VQE algorithm created using Qiskit, which gave results for the ground state energies with an added finite correction. These results match those we are able to find using classical methods, with minimum percent difference. We were able to maintain this accuracy for chiral fermions using the method of Bergman and Thorn. 
In previous runs we found that the amount of Pauli terms grows for larger lattices, particularly those with above 10 lattice sites. Thus this study capped the amount of lattice sites at $n=8$ to  optimize run times. We hope to be able to work with larger lattices in future research. \\

\section*{Acknowledgements}
This material is based upon work supported in part by the U.S. Department of Energy, Office of Science, National Quantum Information Science Research Centers, Co-design Center for Quantum Advantage (C2QA) under contract number DE-SC0012704.This project was supported in part by the U.S. Department of Energy, Office of Science, Office of Workforce Development for Teachers and Scientists (WDTS) under the Science Undergraduate Laboratory Internships Program (SULI).

\end{document}